\documentclass[a4paper,notitlepage,12pt]{article}

\begin{document}

\title{Generalization of the JTZ model to open plane wakes}
\author{ Zuo-Bing Wu\footnotemark[1] \\
State Key Laboratory of Nonlinear Mechanics, \\
Institute of Mechanics, Chinese Academy of Sciences, \\
Beijing 100190, China}

\maketitle

\footnotetext[1]{Author to whom correspondence should be
addressed. Tel: 86-10-82543955; fax: 86-10-82543977. Email:
wuzb@lnm.imech.ac.cn(Z.-B. Wu).}

\newpage
\begin{abstract}
The JTZ model [C. Jung, T. T\'el and E. Ziemniak, Chaos {\bf 3},
(1993) 555], as a theoretical model of a plane wake behind a
circular cylinder in a narrow channel at a moderate Reynolds
number, has previously been employed to analyze phenomena of
chaotic scattering. It is extended here to describe an open plane
wake without the confined narrow channel by incorporating a double
row of shedding vortices into the intermediate and far wake. The
extended JTZ model is found in qualitative agreement with both
direct numerical simulations and experimental results in
describing streamlines and vorticity contours. To further validate
its applications to particle transport processes, the interaction
between small spherical particles and vortices in an extended JTZ
model flow is studied. It is shown that the particle size has
significant influences on the features of particle trajectories,
which have two characteristic patterns: one is rotating around the
vortex centers and the other accumulating in the exterior of
vortices. Numerical results based on the extended JTZ model are
found in qualitative agreement with experimental ones in the
normal range of particle sizes.
\end{abstract}

\textbf{PACS numer(s)}: 05.45.-a, 47.32.-y


\newpage

\textbf{The plane wake behind a circular cylinder is one of the
most fundamental phenomena in fluid mechanics, involving periodic
vortex shedding from the cylinder, which is known as von
K\'arm\'an vortex street. Recently, there emerge vast interests in
the chaotic advection of particles in the wake flow, in particular
in its diversified applications to such processes as chemical and
biological ones. The JTZ model was introduced to describe the
plane wake in a narrow channel at a moderate Reynolds number and
used to investigate chaotic advection of particles near the
circular cylinder. In this paper we extend the JTZ model to
describe an open plane wake without the confined narrow channel by
adding a double row of shedding vortices in the intermediate and
far wake, where the essential change is that large damping in the
original JTZ model is now replaced by little damping in the
extended one. The results of extended JTZ model agree
qualitatively with both direct numerical simulations and
experimental investigations on streamlines and vorticity contours.
By using the extended JTZ model, particle transport processes are
also simulated on different particle sizes. It is shown that the
qualitative features of particle trajectories in the experimental
investigation can be predicted by the numerical simulation,
justifying the application of extended JTZ model to quick and
convenient estimations of the qualitative feature of particle
transports in open plane wake flows.}

\section{Introduction}
The motion of particles in a non-uniform flow has received great
attention due to its potential applications to both natural and
engineering systems[1], as well as its dominant role played in the
transport processes of particulate and multi-phase systems[2],
such as those found in environmental engineering, geophysical
sciences, microfluids and combustion. In the transport processes,
particles are not just passively carried by the background flow,
but also have dynamics of their own. The motion of particles
exhibits abundant characteristics even if the background flows are
very simple. Particles tend to concentrate asymptotically along
periodic, quasi-periodic or chaotic trajectories for such flows as
steady and unsteady cellular flows[3,4], periodic Stuart vortex
flows[5], the von K\'arm\'an vortex street flows[6,7] and plane
wake flows confined in a narrow channel (the JTZ model)[8,9].

The classical von K\'arm\'an vortex street is a well known
pattern[10,11], and the viscous plane wake behind a circular
cylinder, as an example of complex flow containing vortices and
shear layers, has been extensively studied. For exhibiting the
phenomenon of chaotic scattering, a theoretical model (the JTZ
model) was proposed in [12] to describe plane wakes in narrow
channels at a moderate Reynolds number (Re=250). It fits well
 the wake flow field given by direct numerical calculations and
its application is much more convenient in getting the background
wake flow field than solving the Navier-Stokes equations. The
chaotic advection of particles near a circular cylinder has been
investigated by using the JTZ model[13-17], however, relevant
experiments of particle dispersion were conducted in open plane
wakes consisting of near, intermediate and far regions[18,19]. And
thus, we'll extend in this paper the JTZ model to an open plane
wake without the confined narrow channel.

\section{Direct numerical simulation of a plane wake}
 The incompressible plane wake is governed by the two-dimensional Navier-Stokes
equations, as the following non-dimensionalized ones
\begin{equation}
\begin{array}{l}
\nabla \bullet {\bf u}=0,\\
\frac{\partial {\bf u}}{\partial t} = -\nabla {\bf p} + N({\bf u})
+ \frac{1}{Re} L({\bf u}),
\end{array}
\end{equation}
where ${\bf u}$ is the fluid velocity and $p$ is the fluid
pressure divided by the fluid density. The uniform flow velocity
$U_\infty$ and the diameter of circular cylinder $D$ are taken as
the characteristic variables of the system, and Re($=U_\infty D/
\nu_f$, $\nu_f$ is the fluids kinematic viscosity)
 is Reynolds number of
the plane wake. In Eq. (1), the linear diffusion and nonlinear
advection terms are described by
\begin{equation}
\begin{array}{l}
L({\bf u}) = \nabla^2 {\bf u},\\
N({\bf u}) = -\frac{1}{2} [{\bf u} \bullet \nabla {\bf u} + \nabla
\bullet {\bf uu}]. \label{eq4}
\end{array}
\end{equation}
To solve the Navier-Stokes equations, we incorporate a high-order
splitting algorithm into the spectral element method, please refer
to \cite{20,21} for detail. Integrating the Navier-Stokes
equations under the boundary conditions given in \cite{21}, we
obtain velocity field ${\bf u}$ and vorticity field $\omega_z$ at
Re=250. The drag coefficient $C_d$ and Strouhal number $Sr$ are
determined as 1.5 and 0.21, respectively, very close to the
calculation results of \cite{22,23}.

The streamlines and vorticity contours are displayed in Fig. 1,
where a wave structure and the von K\'arm\'an vortex street are
shown in the downstream wake with $x$ restricted to the range of
$x < 20$ to have a clear and regular vortex structure. Since the
wake is a periodic process with a period of $T_c=1/Sr=4.8$, only
figures at time phases 0 and $T_c/4$ in the period $T_c$ are
plotted in Fig.1(a)(b) and Fig.1(c)(d), respectively. The
corresponding figures at time phases $T_c/2$ and $3T_c/4$ can be
obtained by taking mirror images of the plots of Fig.1(a)(b) and
Fig.1(c)(d) about the $x$ axis, respectively. It is observed that
only two vortices can be discerned in the streamline plot of near
wake. The regular von K\'arm\'an vortex street appears in the
intermediate and far wake as shown in the vorticity contour plots,
where the semimajor axes of the elliptical vortices are in the
vertical direction. Close to the cylinder, the distance between
the vortices is about one quarter of those further away. Due to
viscous decay, the vorticity decreases gradually in the wake
downstream. The maximum absolute value of vorticity is very large
in the boundary layer, and drops sharply in the near wake, and
then, the maximum decreases slowly in the intermediate and far
wake,  having two constants in the range of $2<x<10$ and
$10<x<20$, respectively. The maximum in the near wake is about 1.7
times that in the intermediate wake of $2<x<10$, which is about
1.5 times that in the far wake of $10<x<20$. In other words, the
vortex structure in the wake downstream can be captured in a
snapshot and described approximately as follows. Within the wake
region $x < 20$, there occur two transition zones with varying
vorticity and two constant vorticity zones. The first transition
zone is the near wake containing two vortices, and adjacent to
which is the first constant vorticity zone, i.e. the intermediate
wake with four vortices. And further downstream there occurs the
second transition zone containing two vortices, and finally comes
the second constant vorticity zone-the far wake with two vortices.
An experimental investigation on open plane wakes at Re=250 was
given in \cite{24} with time phase around $T_c/2$, the streamline
of which are in close agreement with the above-given simulation
results.

\section{An extended JTZ model}

The classical von K\'arm\'an vortex street is the most simple
model for a plane wake behind a circular cylinder, however, its
computational zone doesn't contain the circular cylinder. The JTZ
model for plane wakes does cover the circular cylinder, and thus
the flow field satisfies the no-slip boundary conditions at the
cylinder surface. In what follows, we give a brief presentation of
the JTZ model, where a plane wake behind a circular cylinder is
considered within a narrow channel at a moderate Reynolds number
(Re=250) with no-slip boundary conditions on the cylinder surface.
By taking the cylinder radius $R_0$ and the vortex shedding period
$T_c$ as the characteristic length and time of the flow, the
dimensionless model stream function[12] is written as

\begin{equation}
\Psi(x,y,t) =f(x,y) g(x,y,t),
\end{equation}
where the first factor
\begin{equation}
f(x,y)=1-{\rm exp}[-a(\sqrt{x^2+y^2}-1)^2]
\end{equation}
satisfies the no-slip boundary conditions at the cylinder surface.
The second factor in Eq. (3) is
\begin{equation}
g(x,y,t) =-wh_1(t) g_1(x,y,t) +wh_2(t) g_2(x,y,t) +u_0ys(x,y).
\end{equation}
The first two terms model the periodic detachment of the vortices,
in which $w$ represents average strength of the time-dependent
vortices and
\begin{equation}
\begin{array}{l}
h_1(t)=|{\rm sin}\pi t|, \\
h_2(t)=h_1(t-\frac{1}{2}).
\end{array}
\end{equation}
The factors
\begin{equation}
\begin{array}{l}
g_1(x,y,t)={\rm exp}^{-\beta_0[(x-x_1(t))^2+\alpha^2 (y-y_0)^2]},\\
g_2(x,y,t)={\rm exp}^{-\beta_0[(x-x_2(t))^2+\alpha^2 (y+y_0)^2]}
\end{array}
\end{equation}
are the Gaussian forms with dimensionless vortex size
$\beta_0^{1/2}$ with its center located at $[x_1(t),y_0]$ and
$[x_2(t),-y_0]$ in the wake. The vortices move downstream at a
constant velocity
\begin{equation}
\begin{array}{l}
x_1(t) =1 +L_0 {\rm mod}(t,1),\\
x_2(t) =x_1(t-\frac{1}{2}).
\end{array}
\end{equation}
 $y_0$ is the distance of the vortex centers from the $x$-axis, $L_0$ is
  the dimensionless distance a vortex traverses during its lifetime and
$u_0$  is the dimensionless background velocity. The last term in
Eq. (5) arises from the background flow, and the screening factor
\begin{equation}
s(x,y) =1-{\rm exp}[-(x-1)^2/\alpha^2 -y^2]
\end{equation}
ensures that the effect of the background flow velocity $u_0$  is
reduced in the wake. In the numerical simulation, parameters are
chosen as $a=1$, $\alpha=2$, $\beta_0=0.35$, $L_0=2$, $y_0= 0.3$,
$u_0= 14$ and $w=24$\cite{15}. The streamlines and vorticity
contours given by the JTZ model at time phases $0$ and $1/4$
in a period are shown in Fig. 2, where the size is rescaled by 1/2
to set the length unit to $D$.

Both the streamlines and vorticity plots show that the JTZ model
gives only two vortices near the circular cylinder, which emerge
and diminish in a period. So the JTZ model fits well the flow
structures of plane wakes in a narrow channel. However, for open
plane wakes without the confined narrow channel, the emerged
vortices won't die out so quickly, and thus it can represent the
flow structures in the near wake, rather than the structures of
staggered double row vortices in the intermediate and far wake. To
extend the JTZ model to the open plane wake, we will preserve the
JTZ model for the near wake and add a double row of staggered
vortices for the intermediate and far wake.

 In the extended stream function  $\Psi(x,y,t)
=f(x,y)g(x,y,t)$, the boundary function $f(x,y)$ is kept, but
$g(x,y)$ is replaced by
\begin{equation}
g(x,y,t) =\gamma_0 \sum_{i=1}^n(-1)^i h_i(t) g_i(x,y,t) +u_0 y
s(x,y),
\end{equation}
where
\begin{equation}
\begin{array}{l}
\gamma_0=\Gamma T_c/R^2_0,\\
u_0=U_\infty T_c/R_0,\\
 g_i(x,y,t) ={\rm exp}^{-\beta_0\{[x-x_i(t)]^2+\alpha^2_i
[y-(-1)^{i-1}y_0]^2\}}, \ \ \ i=1,2,\cdots,n.
\end{array}
\end{equation}
Here,  $n-2$ vortices are added in the intermediate and far wake
flow field, and thus the extended model is more reasonable in
representing the realistic structure of open plane wakes. $\Gamma$
and $U_\infty$ are respectively the magnitude of vortex and
free-stream velocity, and $\gamma_0$ and $u_0$ are their
non-dimensionalized quantities based on $R_0$ and $T_c$.  The
reference velocity is $R_0/T_c$ instead of $U_\infty$, and the
dimensionless inflow velocity is $(u_0,0)$ instead of $(1,0)$. In
Eq. (10), $w$ of the original JTZ model is replaced by $\gamma_0$.
The strength of vortices is defined as time-dependent in view of
the viscous decay mentioned in section 2, where $n(=10)$ vortices
are observed in the range $-2 < x < 20$

\begin{equation}
\begin{array}{l}
\left \{
\begin{array}{l}
h_1=|{\rm sin}\pi t|,\\
h_2=1-0.8{\rm mod}(t,1),\\
h_{3 \le i \le 6}=0.6,\\
h_{7}=0.6-0.2{\rm mod}(t,1),\\
h_{8}=0.5-0.2{\rm mod}(t,1),\\
h_{9 \le i \le 10}=0.4,
      \end{array} \right. {\rm if\ mod}(t,1)<0.5;\\
\left \{
\begin{array}{l}
h_1=1-0.8{\rm mod}(t+0.5,1),\\
h_2=|{\rm sin}\pi (t+0.5)|=|{\rm cos}\pi t|,\\
h_{3 \le i \le 6}=0.6,\\
h_{7}=0.5-0.2{\rm mod}(t+0.5,1),\\
h_{8}=0.6-0.2{\rm mod}(t+0.5,1),\\
h_{9 \le i \le 10}=0.4,
      \end{array} \right. {\rm if\ mod}(t,1) \ge 0.5.
\end{array}
\end{equation}
In the first half of the shedding period, the time dependent
strength of a shedding vortex starts from 0 and reaches 1. In the
second half, it decreases from 1 to 0.6. In the later periods, it
is changing or keeps unchanged depending on its downstream
distance. The vortex centers are located at positions $[x_i(t),
\pm y_0]$, and move downstream with two different velocities
$L_0/2$ and $2L_0$ in the first shedding period and the later
periods, respectively.
\begin{equation}
\begin{array}{l}
x_1(t) =1 +\frac{L_0}{2}{\rm mod}(t,1), \\
x_2(t) =1 +\frac{L_0}{2}{\rm mod}(t+0.5,1),\\
x_i(t) = \left\{ \begin{array}{ll}
        1 +\frac{L_0}{2}+(i-3)L_0 +2L_0 {\rm mod}(t,1),& {\rm if}\ i=2k-1(k \ge 2),\\
        1 +\frac{L_0}{2}+(i-4)L_0 +2L_0 {\rm mod}(t+0.5,1),& {\rm if}\ i=2k(k \ge 2),
                 \end{array}
         \right.
\end{array}
\end{equation}
where $L_0/4$ and $L_0$ are the streamwise distance between two
neighboring vortices within and outside the near wake region,
respectively. Moreover, $s(x,y)$ in the last term of Eq. (10) is
written as follows
\begin{equation}
s(x,y) =1-{\rm exp}[-(x-1)^2/\alpha_0^2 -y^2].
\end{equation}

In order to clarify essential features of the extended JTZ model,
we display in Table 1 the time-variation of the strength and
positions of four vortices near the cylinder in a period. In
general, the up/down vortices in the double row of staggered
vortices are denoted by odd/even. Each pair of vortices exists in
a periodic streamwise interval $[x_{i(=1,3,...)}(t),x_{i+2}(t))$,
which is defined at time phase 0.  For example, at time phase
$0/\frac{1}{2}$, vortex $i=1$ and vortex $i=2$ live in the first
periodic streamwise interval between 1 and $1+L_0/2$, vortex $i=3$
and vortex $i=4$ live in the second one between $1+L_0/2$ and
$1+5L_0/2$, and so on. After one half of the period, i.e., at time
phase $\frac{1}{2}/0$, a new down/up vortex is shed from the
circular cylinder and denoted by 2/1. At the same time, the
down/up vortices in the double row of vortices move away from the
original periodic streamwise positions and are re-located at the
next periodic streamwise positions. Therefore, the notation $i$
for the original down/up row of vortices at time phase
$0/\frac{1}{2}$ will be rewritten as $i+2$. For example, at
$t=1/2$, vortex $i=2$ and vortex $i=4$ are located at $x=1+L_0/2$
and $1+5L_0/2$, respectively. $i=4$ and $i=6$ will be used to
denote the vortex $i=2$ at $x=1+L_0/2$ and vortex $i=4$ at
$x=1+5L_0/2$, respectively. At the same time, a new vortex is shed
at $(1,-y_0)$. It will be denoted by $i=2$. At $t=1$, vortex $i=1$
and vortex $i=3$ are located at $x=1+L_0/2$ and $1+5L_0/2$,
respectively. $i=3$ and $i=5$ will be used to denote the vortices
$i=1$ at $x=1+L_0/2$ and $i=3$ at $x=1+5L_0/2$, respectively. At
the same time, a new vortex is shed at $(1,y_0)$. It will be
denoted by $i=1$. The vortex shedding process is periodic with
period $t=1$.

Parameters $n=10$, $a=1$, $\alpha_0=2$, $\beta_0=0.35$ and $y_0=
0.3$ are used in the following simulations and parameter
$\alpha_i$ is defined as follows

\begin{equation}
\alpha_i= \left \{
\begin{array}{l}
2-1.2{\rm mod}(t,1), \ {\rm if}\ i=1;\\
2-1.2{\rm mod}(t+0.5,1), \ {\rm if}\ i=2;\\
0.8, \ {\rm if}\ i \ge 3,
\end{array} \right.\\
\end{equation}
which describes the continuous change of the horizontal axes of
elliptical vortices from the semimajor in the near wake to the
semiminor in the intermediate and far wake. From the definition of
$Sr$($=2R_0 /U_\infty T_c$), the non-dimensional velocity $u_0$ is
written as $u_0=2/Sr$. For $Re=250$, $Sr$ equals 0.21 as given in
[22]. Based on the velocity and vorticity fields given by the
direct numerical simulation described in section 2, we can
determine approximately $\gamma_0=2.2 u_0$ and $L_0=4.4$.

The streamlines and vorticity contours
  at time phases 0 and $1/4$ in a period are shown in Fig. 3,
  which is better than the original JTZ model in
capturing the global distribution of vortices in the open plane
wake field given by the direct numerical simulation. In
particular, the vorticity distribution in the intermediate and far
wake of the extended JTZ model is found in quantitative agreement
with that of direct numerical simulation, except for the region
near the cylinder surface where there exists some difference. In
the direct numerical simulation, the vorticity is confined to a
thin boundary layer, whereas in the extended JTZ model, it is
extended to a finite domain adjacent to the cylinder surface.
Anyway, the extended JTZ model can be conveniently used as a
simple plane wake model to investigate particle dynamics in plane
wake flows.

\section{Particle dynamics in the extended JTZ model}

For the motion of a small spherical particle in fluid flow, it is
well-known that the flow field around the particle can be
approximated by a Stokes flow provided the particle diameter is
small enough compared to the characteristic length of the fluid
motion. Therefore, the particle-particle interactions as well as
the effects of particles on the flow are negligible. Under these
assumptions, the momentum equation of the motion of the small
spherical particle[25] is written as follows

\begin{equation}
\begin{array}{ll}
\frac{\pi}{6} d_p^3  (\rho_p+0.5 \rho_f) \frac{d{\bf V}}{dt}&=
\frac{\pi}{6} d_p^3  (\rho_p- \rho_f){\bf g}
+\frac{\pi}{4} d_p^3  \rho_f \frac{D{\bf u}}{Dt}\\
&+3 \pi d_p \nu_f \rho_f ({\bf u}-{\bf V}) \nonumber + \frac{3}{2}
(\pi \nu_f)^{1/2} d_p^2 \rho_f \int_0^t \frac{1}{\sqrt{t-\tau}}
(\frac{d{\bf u}}{d\tau}-\frac{d{\bf V}}{d\tau}) d\tau,
\end{array}
\end{equation}
where ${\bf V}$ is velocity of the particle, and $\rho$ is
density, ${\bf g}$ is gravitational acceleration, and subscripts
$f$ and $p$ refer to the fluid and particle, respectively.  The
derivatives $D/Dt$ and $d/dt$ are used to denote the time
derivative following a fluid element and the moving sphere,
respectively. Introducing the dimensionless quantities
$\delta=\rho_p / \rho_f$, $\epsilon=1/(0.5+\delta)$, ${\bf
x}^*={\bf x}/R_0$, $t^*=t/(R_0/U_\infty)$, ${\bf u^*}={\bf
u}/U_\infty$, ${\bf V^*}={\bf V}/U_\infty$ and ${\bf g^*}={\bf
g}/g$, based on $R_0$ and $U_\infty$, we can non-dimensionlize
Eq.(16) as follows

\begin{equation}
\begin{array}{ll}
\frac{d{\bf V}}{dt}&= \frac{(1-1.5\epsilon)}{Fr^2} {\bf
g}+\frac{3\epsilon}{2} \frac{D{\bf u}}{Dt} +\frac{1}{St} ({\bf
u-V}) \\
&+3 \sqrt{\frac{\epsilon}{2 \pi St}} \int_0^t
\frac{1}{\sqrt{t-\tau}} (\frac{d{\bf u}}{d\tau}-\frac{d{\bf
V}}{d\tau}) d\tau,
\end{array}
\end{equation}
where  $Fr=U_\infty /\sqrt{gR_0}$ and $St=U_\infty T/R_0$ ($T$ is
the particle viscous relaxation time, $d^2_p/18 \epsilon \nu_f$)
are the Froude number and the Stokes number, respectively. The
density and size of particles in Eq.(16) are taken from [18] as
$\rho_p=2.4 \times 10^3 kg/m^3$ and $d_p=0(10^{-5})m$,
respectively. Since air is chosen as the fluid medium in the flow,
the fluid properties in Eq.(16) are described by $\rho_f=1.225
 kg/m^3$ and $\nu_f=1.45 \times 10^{-5}m^2/s$[26]. To emphasize
 the effects of viscous force or $St$ in Eq. (17) on particle transport
 process, the background flow field parameters are chosen as
 $U_\infty=0.1m/s$ and $R_0=1.81 \times 10^{-2}m$.

Since $\delta$ is fixed and is of the order of $0(10^3)$,
parameter $\epsilon$ appears to be of the order of $0(10^{-3})$.
The particle diameters $d_p$ is very small [$d_p=0(10^{-5}) m$],
so the parameters $1/St$ and $1/Fr^2$ appear to be of the order of
$0(10^1)$ and $0(10^0)$, respectively. In this case, we can
neglect the following smaller order terms: the gravity term, the
stress tensor term, the Basset history term and reduce the
equation of motion (17) to
\begin{equation}
\frac{d{\bf V}}{dt}= \frac{1}{St} ({\bf u-V}).
\end{equation}
 In section 3, the
physical quantities are non-dimensionalized using $T_c$ and $R_0$,
which are different from the above ones. The function ${\bf
u}(x,y,t)$ in Eq. (18) is now replaced by
\begin{equation}
\begin{array}{l}
{\bf u}(x,y,t) =\hat{\bf u}(x,y,\hat{t})/u_0,\\
\hat{t}=\frac{Sr}{2} t,
\end{array}
\end{equation}
where $\hat{t}$ and $\hat{\bf u}$ are the time and velocity field
for the extended JTZ model, respectively. In the
definition of $Sr$, the characteristic length is 2$R_0$, rather
than $R_0$, which is the characteristic length of the model. Since
$\hat{t} \times T_c = t \times R_0/U_\infty$, so $\hat{t} = t
\times R_0/(U_\infty T_c) = t \times Sr/2$. In what follows, we
will analyze essential features of particle trajectories in the
extended JTZ model flow.

Equation (19) is integrated with time step $\Delta t=0.001-0.01$
by using a fourth-order Runge-Kutta algorithm. Particles are
released along a semicircle (of radius 1.2) with time interval $0.05$.
In each time interval, the released particles total to
100. The initial velocities of particles are taken as the local
flow ones. We choose $\hat{t}=0$ as the time for the initial
release of particles, and give a snapshot of particle
distribution in the wake at $\hat{t}=3$.

The particle diameter $d_p$ is taken in the range of $8 \times
10^{-6} - 8 \times 10^{-5}$m, which corresponds to $St=0.003 -
0.27$. In the numerical simulation of $d_p=8 \times 10^{-6}$m, $1
\times 10^{-5}$m, $2 \times 10^{-5}$m, $3 \times 10^{-5}$m, $5
\times 10^{-5}$m and $8 \times 10^{-5}$m, the particles rotate
around neighboring vortices as well as move downstream under the
effects of viscous force $({\bf u}- {\bf V })/St$. In interacting
with vortices, they exhibit some characteristic distributions.
According to the rotation directions of vortices, the particle
clusters form foliations near the vortices, and show two typical
patterns depending on the particle sizes. Firstly, when the
particle diameter is small enough (in the range of $8 \times
10^{-6}-2 \times 10^{-5}$m), the foliations rotate around the
vortex centers with an example shown in Fig. 4(a) for $d_p=1
\times 10^{-5}$m ($St=0.004$). The large viscous force provides
centripetal force to particles, and thus the foliations are folded
by the rotation of the vortices and stretched by the training
field. When the particle diameter is in the range of $2 \times
10^{-5}-8 \times 10^{-5}$m, the foliations are accumulated in the
exterior of vortices with an example shown in Fig. 4(b) for $d_p=5
\times 10^{-5}$m ($St=0.1$). Since the viscous force decreases as
$d_p$ increases, the larger the $d_p$, the smaller the centripetal
force provided by the viscous force. Hence, the foliations are
moved away from the vortex centers and accumulate in the exterior
of vortices.  In the experiments of [18,19],
 the particle dispersion in a plane wake is investigated in detail.
 As the particle size increases, the organized pattern of particle
 trajectories changes from filling in the vortex structures to
 accumulating along the exterior of vortices. In the direct numerical
simulation of [27], the particle distributions for different
particle sizes are similar to the above-mentioned experimental
observations, and these behaviors agree qualitatively with our
numerical results based on the extended JTZ model.

\section{Conclusion}
In summary, we generalize and extende the JTZ model, which is a
theoretical model for a plane wake behind a circular cylinder in a
narrow channel at a moderate Reynolds number, to describe an open
plane wake without the confined narrow channel by incorporating a
global distribution of shedding vortices. The extended JTZ model
is found in qualitative agreement with both direct numerical
simulations and experimental studies in respect of streamlines and
vorticity contours. The interaction of small spherical particles
with vortices in the extended JTZ model flow is investigated to
further validate the moderling in studying particle transport
processes. It is found that the particle size has significant
influences on the features of particle trajectories, and the
particles exhibit two typical patterns as the particle size
increases. When the particle size is small enough, particle
clusters would rotate around the vortex centers, whereas they
would accumulate in the exterior of vortices for larger particle
size. The numerical results are found in qualitatively agreement
with the experimental ones in the normal range of particle sizes.

\textbf{Acknowledgments} The author thanks ICTS and IMECH research
computing facilities for assisting in the computation.

\newpage
Table I. Time-variation of the strength and positions of four
vortices near the cylinder in a period.

\begin{tabular}{l|llll}
 \hline
t          &  0         & 1/4          & 1/2       & 3/4\\
\hline
$h_1$      &  0         & $\sqrt{2}$/2 &1          & 0.8 \\
$h_2$      &  1        & 0.8            &0          & $\sqrt{2}$/2 \\
$h_3$      &  0.6         & 0.6            &0.6          &  0.6\\
$h_4$      &  0.6         & 0.6            &0.6          &  0.6\\
$x_1$      &  1         & 1+$L_0$/8    &1+$L_0/4$  & 1+3$L_0$/8 \\
$x_2$      &  1+$L_0$/4 & 1+3$L_0$/8   &1          & 1+$L_0$/8 \\
$x_3$      &  1+$L_0$/2 & 1+$L_0$      &1+3$L_0$/2 & 1+2$L_0$ \\
$x_4$      &  1+3$L_0$/2& 1+2$L_0$     &1+$L_0$/2  & 1+$L_0$\\
 \hline
\end{tabular}

\newpage
\textbf{FIGURE CAPTION}

Fig. 1. Streamlines and vorticity contours for the plane wake flow
at $Re=250$ obtained from direct numerical simulation. The values
of vorticity $\omega_{D, U_\infty}$, which are non-dimensionalized
by $D$ and $U_\infty$, are shown in the vorticity contour plots.
The time phases are chosen as $0$ in (a)$-2 < x < 10; $(b) $10 < x
< 20$ and $T_c/4$ in (c)$-2 < x < 10$; (d)$10 < x < 20 $ in a
vortex shedding period $T_c$.

Fig. 2. Streamlines and vorticity contours for the JTZ model
flow at time phases (a) $0$ and (b) $1/4$ in a period.

Fig. 3. Streamlines and vorticity contours for the extended JTZ
model flow at time phases 0 in (a)$-2 < x < 10; $(b) $10 < x < 20$
and $1/4$ in (c)$-2 < x < 10$; (d)$10 < x < 20 $ in a period. The
index $i$ of vortices in the extended JTZ model (10) is marked at
the upper and bottom zones of the vortex street central region.
The non-dimensional values $\omega_{D, U_\infty}$ in the vorticity
contour plots are calculated from the modelling of vorticity
$\omega_{R_0,T_c}$ scaled by $R_0$ and $T_c$ by referring to the
relation $\omega_{D,U_\infty}=Sr
\omega_{R_0,T_c}=0.21\omega_{R_0,T_c}$.

Fig. 4. Instantaneous particle distribution and vorticity contour
in the extended JTZ model flow at time phase $0$ in a
period with different particle diameters (a) $d_p=1 \times
  10^{-5}$m and (b) $d_p=5 \times 10^{-5}$m. The release of particles
  starts at $\hat{t}=0$ and is completed at $\hat{t}=3$ when the snapshot
  is taken.


\newpage
\begin{thebibliography}{}

\bibitem{1}
R. Clift, J.R. Grace and M.E. Weber, {\it Bubbles, Drops and
Particles}, Academic, NY, 1978.
\bibitem{2}
D. De Kee and R. P.Chhabra, {\it Transport Processes in Bubbles,
Drops and Particles}, Taylor \& Francis, NY, 2002.
\bibitem{3} L. P. Wang, M. R. Maxey, T. D. Burton and D. E. Stock,
Chaotic dynamics of particle dispersion in fluids, {\it Phys,
Fluids} A4 (1992) 1789.
\bibitem{4} J. C. Zahnow and U. Feudel,
Moving finite-size particles in a flow: A physical example of
pitchfork bifurcations of tori, {\it Phys. Rev.} E77 (2008)
026215.
\bibitem{5}
K.-K. Tio, A.M. Ganan-Calvo and J.C. Lasheeras, The dynamics of
small, heavy, rigid spherical particles in a periodic Stuart
vortex flow, {\it Phys. Fluids} A5, (1993) 1679.
\bibitem{6}
Z.-B. Wu, G.-C. Ling,  and Q. J. Xing, Effects of particle size on
dilute particle dispersion in a Karman vortex street flow, {\it
Chin. Phys. Letts.} 29, (2002) 83.
\bibitem{7}
Z.-B. Wu, Streamline topology and dilute particle dynamics in a
Karman vortex street flow, {\it Int. J. Bifur. and Chaos} 13,
(2003) 1275.
\bibitem{8}
I.J. Benczik, Z. Toroczkai and T. T\'el, Selective sensitivity of
open chaotic flows on inertial tracer advection: catching
particles with a stick, {\it Phys. Rev. Letts.} 89, (2002) 164501.
\bibitem{9}
 I. J. Benczik, Z. Toroczkai and T. T\'el, Advection of
 finite-size particles in open flows,
 {\it Phys. Rev. E} 67, (2003) 036303.
\bibitem{10}
G. Lamb, {\it Hydrodynamics}, 6th ed. Cambridge Univ. Press,
London, 1978.
\bibitem{11}
P. G. Saffman, {\it Vortex dynamics}, Cambridge Univ. Press,
London, 1992.
\bibitem{12}
C. Jung, T. T\'el and E. Ziemniak, Application of scattering chaos
to particle transport in a hydrodynamical flow, {\it Chaos} 3,
(1993) 555.
\bibitem{13}
E. Ziemniak, C. Jung and T. T\'el, Tracer dynamics in open
hydrodynamical flows as chaotic scattering, {\it Physica D} 76,
(1994) 123.
\bibitem{14}
Y. Do and Y. C. Lai, Stability of attractors formed by inertial
particles in open chaotic flows, {\it Phys. Rev. E} 70, (2004)
036203.
\bibitem{15}
T. T\'el, A. Moura, C. Grebogi and G. K\'arolyi, Chemical and
biological activity in open flows: A dynamical system approch,
{\it Phys. Reports} 413, (2005) 91.
\bibitem{16}
M. Sandulescu, E. Hernandez-Garcia, C. Lopez and U. Feudel,
Kinematic studies of transport across an island wake, with
application to the Canary islands, {\it Tellus A} 58, (2006) 605.
\bibitem{17}
M. Sandulescu, C. Lopez, E. Hernandez-Garcia and U. Feudel,
Plankton blooms in vortices: the role of biological and
hydrodynamic timescales, {\it Nonlin. Proc. Geophys.} 14, (2007)
443.
\bibitem{18}
L. Tang, F. Wen, Y. Yang, C.T. Crowe, J.N. Chung and T.R. Troutt,
Self-organizing particle dispersion mechanism in a plane wake,
{\it Phys. Fluids} A4, (1992) 2244.
\bibitem{19} C.T. Crowe, J.N. Chung and T.R. Troutt, Particle interaction with
vortices. in {\it Fluid Vortices}, ed. Green SI, Kluwer Academic
Publishers: Dordrecht, 1995; 829-858.
\bibitem{20}
G. E. Karniadakis and S. J. Sherwin, {\it Spectral/hp element
methods for CFD}, Oxford University Press, New York, 1999.
\bibitem{21}
Z.-B. Wu, Numerical study on heavy rigid particle motion of a
plane wake flow by spectral element method, {\it Int. J. Num.
Method in Fluids} 61, (2009) 536.
\bibitem{22}
H.Q. Zhang, U. Fey and B. R. Noack, On the transition of the
cylinder wake, {\it Phys. Fluids} 7, (1995) 779.
\bibitem{23}
D. Barkley and R. D. Henderson, Three-dimensional Floquest
stability analysis of the wake of a circular cylinder, {\it J.
Fluid Mech.} 322, (1996) 215.
\bibitem{24}
L.M. Milne-Thomson, {\it Theoretical hydrodynamics}, 5th ed. The
Macmillan Press, London, 1979.
\bibitem{25} M.R. Maxey and J.J. Riley,  Equation of motion for a small rigid sphere
in a nonuniform flow. {\it Phys. Fluids}  26, (1983) 883.
\bibitem{26}
R. L. Panton, {\it Incompressible Flow}, John Wiley \& Sons: New
York, 1984.
\bibitem{27}
A. M. Froncioni, F. J. Muzzio, R. L. Peskin and P. D. Swanson,
Chaotic motion of fluid and solid particles in plane wakes. {\it
Chaos, Solitions \& Fractals} 8, (1997) 109.

\end{thebibliography}
\end{document}